\documentclass[11pt] {article}
\usepackage{graphicx}       
\usepackage{bm}         
\usepackage{amsfonts}
\usepackage{amsmath}        
\usepackage{latexsym}

\begin{document}

\title{Transformation design and nonlinear Hamiltonians}
\author{Thomas Brougham$^{\ast}$$\thanks{$^\ast$ Corresponding author.  Email: thomas.brougham@gmail.com}$, Goce Chadzitaskos and Igor Jex\\
Department of Physics, FNSPE, Czech Technical University in Prague,
B\v{r}ehov\'{a} 7,\\ 115 19 Praha 1, Czech Republic.} \maketitle
\abstract{We study a class of nonlinear Hamiltonians, with
applications in quantum optics. The interaction terms of these
Hamiltonians are generated by taking a linear combination of powers
of a simple `beam splitter' Hamiltonian.  The entanglement
properties of the eigenstates are studied. Finally, we show how to
use this class of Hamiltonians to perform special tasks such as
conditional state swapping, which can be used to generate optical
cat states and to sort photons.
\\ keywords: quantum optics, nonlinear optics, quantum information}

\section{Introduction}
The consideration of nonlinear optical processes within quantum optics has led to the study of many important physical
phenomena.  These nonlinear optical processes have found many diverse applications, such as generating entangled photons
using parametric down conversion \cite{para} or creating optical routers using intensity dependent properties of
an optical fibre \cite{jensen, chefles}.  The growing field of quantum information has added to the interest
in nonlinear optics.  This is due to the fact that nonlinear optics can be used to perform certain operations that
are not possible using only linear optics.   For example, a class of nonlinear optical Hamiltonians, which have been
used to model four wave mixers  \cite{milburn,yurke,tombesi, yurke2}, have been shown to allow one to prepare
a macroscopically distinguishable superposition of quantum states.

In this paper we will consider a class of nonlinear Hamiltonians.
The Hamiltonians will be functions (e.g. a polynomial) of a
quadratic Hamiltonian, $\hat H_0$, that leads to linear differential
equations in the field mode operators.  Hamiltonians of this type
have been used to describe optical beam splitters.  The nonlinear
Hamiltonians will thus commute with $\hat H_0$.  Hence the task of
obtaining the eigenvectors and eigenvalues of the nonlinear
Hamiltonians will reduce to diagonalising the simple quadratic
Hamiltonian.  It will be shown that the eigenvalues and eigenvectors
are related to a class of orthogonal Krawtchouk polynomials
\cite{askey}.  We thus establish that this particular class of
nonlinear Hamiltonians are exactly solvable.  The connection between
orthogonal polynomials and nonlinear optical process has been
studied previously and general results have been obtained
\cite{Goce2, Goce,Horow}.  In particular, the mathematical results
that we present in section \ref{sec2} is  a special case of the
general theory presented in \cite{Goce2}.  In \cite{Goce2} the
general method is described for solving two boson systems in quantum
optics via of orthogonal polynomials systems. It is possible to
study many of these problems using different mathematical
techniques.  In particular, beam splitters have bean studied
extensively \cite{klauder, campos, leonhardt}. The benefit of using
orthogonal polynomials, however, is that there exists extensive
literature on their properties, which one can exploited for any
given problem, see for example \cite{askey, orthop}.


The organization of the paper will be as follows.  In section
\ref{sec2} we will diagonalise the quadratic Hamiltonian.  This will
be achieved by first showing the equivalence between diagonalising
the Hamiltonian and solving a particular recurrence relation.  The
solutions of the recurrence relation are the Krawtchouk polynomials.
The properties of these polynomials will be used to obtain the
spectrum of the Hamiltonian.  In section \ref{sec3} we study the
entanglement properties of the energy eigenstates.  The properties
of the nonlinear Hamiltonians will be investigated in section
\ref{sec4}.  In particular we will show how the Hamiltonians can be
used to perform a conditionally swap operation on the two modes,
when the modes are both prepared with all the photons initially in
one mode.  We will also show how this enables one to prepare optical
Schr\"{o}dinger cat states. The dynamics of the nonlinear
Hamiltonians is studied in further detail in section \ref{sec5}.
Finally, we discuss our results in section \ref{conc}.

\section{Diagonalising the quadratic Hamlitonian}
\label{sec2}

In this section we apply the method investigated in \cite{Goce2}.
  Suppose we have an electromagnetic field that has two modes 1 and
2. Associated with these modes are the creation and annihilation
operators $\hat a^{\dagger}_j$ and $\hat a_j$, where $j=1,2$.  These
operators obey the standard canonical commutator relations $[\hat
a_j,\hat a^{\dagger}_j]=\hat 1$ and $[\hat a_j,\hat a_k]=0$, for
$j\ne k$.  Suppose further that the two modes interact.  If this
interaction is quadratic, then one possible form for the interaction
Hamiltonian is
\begin{equation}
\label{bsh}
\hat H_0=\gamma\hat a^{\dagger}_{1}\hat a_{2}+\gamma^*\hat a^{\dagger}_{2}\hat a_{1},
\end{equation}
where the free evolution terms are ignored\footnote{One could imagine that we are working in an interaction picture.}  Hamiltonians of the form (\ref{bsh}) are be used to describe beam splitters.  A simple class of nonlinear interaction Hamiltonians can be constructed by introducing some nonlinear function $f(x)$ and forming a new Hamiltonian $\hat H=f(\hat H_0)$, where $\hat H_0$ is defined in equation (\ref{bsh}).  A simple example of this would be $f(x)$ a  polynomial e.g.
\begin{equation}
\label{nonlinear}
\hat H=\mu\hat H_0+\lambda(\hat H_0)^2.
\end{equation}
The nonlinear Hamiltonian described in equation (\ref{nonlinear} has
been studied before in connection with various nonlinear optical
phenomena, see for example \cite{milburn,tombesi}.

The quadratic Hamiltonian $\hat H_0$ conserves the total number of photons, i.e.
\begin{equation}
[\hat n,\hat H_0]=0,
\end{equation}
where $\hat n=\hat n_1+\hat n_2$.  The class of nonlinear
Hamiltonians $\hat H=f(\hat H_0)$, where $f(x)$ is some polynomial,
will also conserve the total number of photons.  By construction we
see that $[\hat H_0,\hat H]=0$.  We can thus obtain the eigenvectors
of $\hat H$ by diagonalising $\hat H_0$.  The fact that both
Hamiltonians conserve the total photon number suggests that we
should diagonalise the Hamiltonian (\ref{bsh}) in subspaces with
fixed numbers of photons.  If we have a subspace where the total
number of photons is constant and equals $M$, then the dimensions of
this subspace will be $M+1$ and any state belonging to the subspace
can be written in the form
\begin{equation}
|\psi^M\rangle_{12}=\sum_{n=0}^{M}{\xi_n|M-n,n\rangle_{12}},
\end{equation}
where $|M-n,n\rangle_{12}$ represents the Fock state $|M-n\rangle_1\otimes|n\rangle_2$.  Let $|E^M\rangle_{12}$ denote an eigenstate of (\ref{bsh}), hence $\hat H_0|E^M\rangle_{12}=E|E^M\rangle_{12}$.  For the  sake of convenience we shall express the eigenstate in the following form
\begin{equation}
\label{evector}
|E^M\rangle_{12}=\sum^{M}_{n=0}{c^M_ne^{-ing}|M-n,n\rangle_{12}},
\end{equation}
where $g=\arg(\gamma)$, i.e. $\gamma=|\gamma|e^{ig}$.  Acting on equation (\ref{evector}) with $\hat H_0$ yields the expression
\begin{eqnarray}
\hat H_0|E^M\rangle_{12}=\gamma\sum_n{\sqrt{n(M-n+1)}e^{-ing}c^M_{n}|M-n+1,n-1\rangle_{12}}\nonumber\\
+\gamma^*\sum_n{\sqrt{(n+1)(M-n)}e^{-ing}c^M_n|M-n,n+1\rangle_{12}}=E|E^M\rangle_{12}.
\end{eqnarray}
By re-arranging this expression one can show that the coefficients $c^M_n$, obey the following three term recurrence relation
\begin{equation}
\label{3term}
\frac{E}{|\gamma|}c^M_n=\sqrt{(n+1)(M-n)}c^M_{n+1}+\sqrt{n(M-n+1)}c^M_{n-1}.
\end{equation}
The recurrence relation (\ref{3term}) can be solved to find both the eigenvalues and the coefficients that appear in the Fock basis expansion of the energy eigenvectors.  The mathematical properties of the recurrence relation (\ref{3term}) enable us to determine many properties of the eigenvectors and eigenvalues of energy.  For example, if equation (\ref{evector}) is an eigenvector of $\hat H_0$ corresponding to the eigenvalue $E$, then so is the vector $|-E^M\rangle_{12}=\sum_n{(-1)^n\exp(-ing)c^M_n|M-n,n\rangle_{12}}$, which corresponds to the eigenvalue $-E$.  The proof of this fact follows simply by replacing the terms $c^M_n$, in (\ref{3term}), with $(-1)^{n}c^M_n$.  We thus see that for every positive eigenvalue of energy there must exist a corresponding negative energy eigenvalue.

The coefficients $c^M_n$ will be functions of $E$.  If we formally solve the recurrence relation (\ref{3term}), then we find that $c^M_n(E)$ is a polynomial of degree $n$, in the variable $E$.  The fact that the total number of photons is $M$ means that $c^M_{M+1}(E)=0$, which means that the roots of the polynomial $c^M_{M+1}(E)$ correspond to the eigenvalues of (\ref{bsh}).  To help explain the last fact consider the matrix representation of equation (\ref{3term})
\begin{equation}
\frac{E}{|\gamma|}\left(\begin{matrix}
c^M_0\\
c^M_1\\
\vdots\\
c^M_M\end{matrix}
\right)=\left(\begin{matrix}
0 & \sqrt{M} & 0 & 0 & \ldots & 0 \\
\sqrt{M} & 0 & \sqrt{2(M-1)} & 0 & \ldots & 0\\
\vdots & \vdots & \vdots & \vdots  & \vdots & \vdots\\
0 &\ldots &\ldots & \sqrt{M} & 0\end{matrix}
\right)\left(\begin{matrix}
c^M_0\\
c^M_1\\
\vdots\\
c^M_M\end{matrix}\right)
+\left(\begin{matrix}
0\\
0\\
\vdots\\
\phi_{M+1}c^M_{M+1}\end{matrix}\right),
\end{equation}
where $\phi_{M+1}$ is the term that we would find in front of $c^M_{M+1}$ in the recurrence relation (\ref{bsh}).  It can be seen that if $c^M_{M+1}(E)=0$, then the column vector ${\bf c}^M$ will be an eigenvector of the Hamiltonian, corresponding to the eigenvalue $E/|\gamma|$.  An important point to note about equation (\ref{3term}) is that the factors that multiply both $c^M_{n+1}$ and $c^M_{n-1}$ are positive.  This fact enables us to make use of a theorem of Favard's \cite{favard}.  For a recurrence relation of the form
\begin{equation}
xP_n(x)=b_{n+1}P_{n+1}(x)+b_{n-1}P_{n-1}(x),
\end{equation}
where the sequence $\{b_n\}$ is non-negative, there exists a linear functional $\mathcal{L}$, which has the following properties
\begin{eqnarray}
\mathcal{L}[1]&=&1,\nonumber\\\mathcal{L}[P_m(x)P_n(x)]&=&0\;\,\text{for}\;m\ne n,\nonumber\\
\mathcal{L}[P^2_n(x)]&\ne& 0.
\end{eqnarray}
Applying this theorem to (\ref{3term}) leads to the conclusion that
the set of polynomials $\{c^M_n(x)\}$, will be orthogonal to each
other with respect to the linear functional $\mathcal{L}$.  A
standard property of orthogonal polynomials is that the roots of an
orthogonal polynomial are distinct \cite{mathsphys}.  This last
result together with the fact that the eigenvalues of $\hat H_0$ are
the roots of the polynomial $c^M_{M+1}(E)=0$, implies that the
eigenvalues of $\hat H_0$ are nondegenerate, when we are confined to
the $M$ photon subspace\footnote{The spectrum of $\hat H_0$ on the
whole Hilbert space will have degeneracies due to the fact that the
spectrum of $\hat H_0$ on the different subspace is not disjointed
and thus can share eigenvalues.}.  There is an extensive literature
on orthogonal polynomials and their properties
\cite{askey,orthop,mathsphys}.  One important example are the
Krawtchouk polynomials, which obey the following three term
recurrence relation \cite{askey}
\begin{eqnarray}
\label{krawrec}
xK_n(x;p,M)&=&\sqrt{p(1-p)(n+1)(M-n)}K_{n+1}(x;p,M)+\left(p[M-n]+n[1-p]\right)K_{n}(x;p,M)+\nonumber\\
&&+\sqrt{p(1-p)n(M-n+1)}K_{n-1}(x;p,M),
\end{eqnarray}
where $0 < p < 1 $.  If we set $p=1/2$, $2x-M=E/|\gamma|$ and $K_n(x;1/2,M)=c^M_n(E)$, then equation (\ref{krawrec}) reduce to equation (\ref{3term}).  The coefficients $c^M_n(E)$ are thus Krawtchouk polynomials.  The form of the Krawtchouk polynomials, for this choice of parameters, is given by
\begin{eqnarray}
\label{krawpol} K_{n} (x;1/2,M) =
(-1)^nK_0(x;1/2,M)\sqrt{\frac{M!}{n!(M-n)!}}
\sum^{n}_{k=0}\frac{(-n)_k (-x)_k }{(-M)_{k} k!}2^{k},
\end{eqnarray}
where $(y)_{k}= y (y+1)(y+2)\ldots(y+k-1)$.  The roots of the
Krawtchouk polynomials can be determined using (\ref{krawpol}).  It
can easily be shown that the roots of $K_{M+1}(x;1/2,M)=0$ are $x=0,
1, 2, ..., M$, which implies that the energy eigenvalues are
\begin{equation}
\label{eenergy}
E=-M|\gamma|,(2-M)|\gamma|,(4-M)|\gamma|,...,M|\gamma|
\end{equation}
By diagonalising the Hamiltonian (\ref{bsh}) we have solved the system's dynamics and can analyze relevant quantum effects.

\section{Photon distribution and entanglement properties of the energy eigenstates}
\label{sec3} In the previous section we found the analytic solution
of equation (\ref{3term}), which in turn gives us the form of the
eigenvectors of the Hamiltonian (\ref{bsh}).  Nevertheless, it is
informative to give some simple examples of the eigenvectors of
(\ref{bsh}).  For the simplest case, when $M=1$, the eigenvalues are
just $E=\pm|\gamma|$ and the eigenvectors take the form
$|E\pm\rangle=(|10\rangle\pm\exp(-ig)|01\rangle)/\sqrt{2}$.  When
$M=2$, we find that $E=\pm2|\gamma|, 0$ and
\begin{eqnarray}
\label{evecm2}
|E=\pm2|\gamma|\rangle&=&\frac{1}{2}\left(|20\rangle\pm\sqrt{2}e^{-ig}|11\rangle+ e^{-2ig}|02\rangle\right),\nonumber\\
|E=0\rangle&=&\frac{1}{\sqrt{2}}\left(|20\rangle-e^{-2ig}|02\rangle\right).
\end{eqnarray}
As $M$ increases the form of the eigenvectors will generally become more complicated.   One situation, however, where the eigenvectors have a simple form is when $E=M|\gamma|$, i.e for the eigenvectors corresponding to the maximum eigenvalue of $\hat H_0$.  In this case we find
\begin{eqnarray}
\label{ccc}
c^M_n=K_{n} (M;1/2,M)=(-1)^nc^M_0\sqrt{\frac{M!}{n!(M-n)!}}\sum^{n}_{k=0}\frac{(-n)_k }{k!}2^{k}\nonumber\\
=(-1)^nc^M_0\sqrt{\frac{M!}{n!(M-n)!}}\sum^{n}_{k=0}\frac{n!}{k!(n-k)!}(-2)^{k}=c^M_0\sqrt{\frac{M!}{n!(M-n)!}}.
\end{eqnarray}
If $c^M_0$ is chosen to be $2^{-M/2}$, then
$\sum_n{|c^M_n(M|\gamma|)|^2}=1$.  A consequence of this
result is that the photon probability distribution, $|c^M_n|^2$, will be a binomial distribution when $E=M|\gamma|$.  In the limit of large $M$,
$|c^M_n(M|\gamma|)|^2$ can be approximated by a Gaussian. 

\begin{figure}
\center{\includegraphics[width=14cm, height=!]
{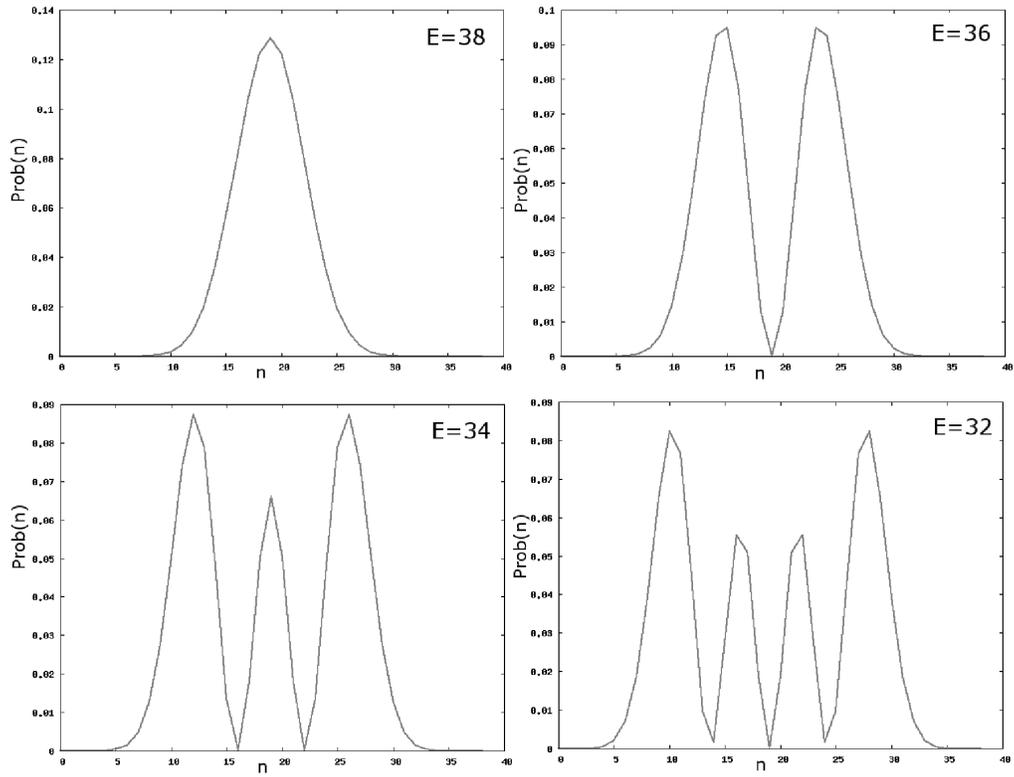}}
\caption{Four different plots of the probability distribution for finding $n$ photons in the second mode.  Each plot is for the situation where $M=38$ and $|\gamma|=1$, however in each successive plot the energy is decreased by $2|\gamma|$.}
\label{fig1}
\end{figure}

The probability distribution $|c^M_n|^2$, of finding $n$ photons in
the second mode, depends only on $|E|$.  By this we mean that
$|c^M_n(E)|^2=|c^M_n(-E)|^2$.  This follows directly from the result
in section \ref{sec2} that $c^M_n(-E)=(-1)^n c^M_n(E)$, which can be
verified using the recurrence relation (\ref{3term}).  The behaviour
of the probability distributions $|c^M_n(E)|^2$, for
$|E|<M|\gamma|$, can be studied by numerically evaluating
(\ref{krawpol}).  Figure \ref{fig1} shows $|c^M_n|^2$ plotted with
$M=38$, for the four largest eigenvalues of $\hat H_0$.  We see that
each time $E$ is decreased by $2|\gamma|$, an additional peak
appears in the probability distribution.  Numerical investigations
show that this behaviour is a generic feature and occurs for each
value of $M$.  Figure \ref{fig1} shows that the probability
distribution has the symmetry, $|c^M_n|^2=|c^M_{M-n}|^2$.  This
means that the probability of finding $n$ photons in the first mode
is the same as finding $n$ photons in the second mode.  This result
is physically reasonable due to the symmetric nature of the
Hamiltonian (\ref{bsh}), with regards to the two modes.


\begin{figure}
\center{\includegraphics[width=8cm, height=!]
{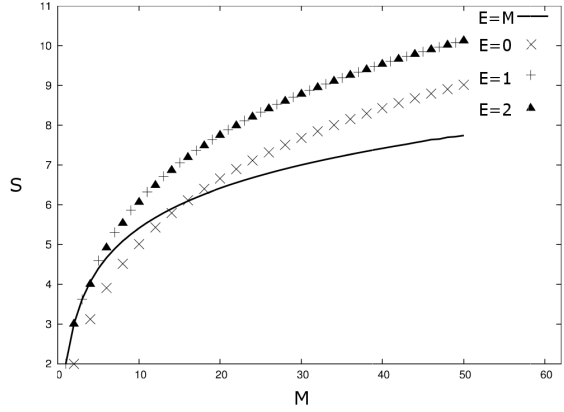}}
\caption{A plot of the entanglement, $S_{ent}$, against the number of photons $M$, for different values of energy $E$, with $\gamma=1$.}
\label{fig3}
\end{figure}

If we calculate the explicit forms for the energy eigenvectors, then we see that the two modes are entangled, i.e they cannot be express in a factorised  form.  Entanglement is one of the defining characteristics of quantum mechanics and has great importance in the foundations of quantum mechanics and in the field quantum information \cite{nch}.  For this reason it is of interest to quantify the entanglement present in the energy eigenstates.  A common approach to quantifying the entanglement in a bipartite pure state is to calculate the von Neumann entropy in the reduced state \cite{nch,steve,steve2}.  The von Neumann entropy of a density operator $\hat \rho$ is defined to be
\begin{equation}
S[\hat\rho]=-\text{Tr}[\hat\rho\log(\hat\rho)],
\end{equation}
where the logarithms are taken to the base 2.  The reduced states of the bipartite state $|\psi\rangle_{12}$ are define to be $\hat\rho_1=\text{Tr}_2[|\psi\rangle\langle\psi|]$ and $\hat\rho_2=\text{Tr}_1[|\psi\rangle\langle\psi|]$, where $\text{Tr}_j$ denotes the partial trace over the subsystem $j$.  We thus take use the following quantity to measure the entanglement \cite{steve}
\begin{equation}
\label{eform}
S_{ent}(|\psi\rangle\langle\psi|)=2S(\hat\rho_1)=2S(\hat\rho_2).
\end{equation}
$S_{ent}(|\psi\rangle\langle\psi|)=0$ when our bipartite state is separable.  In addition to this, we can see that equation (\ref{eform}) assumes its maximum value when our bipartite state is a maximally entangled state, i.e. the reduced states are proportional to the identity operator.  If we express the energy eigenvectors in the form given in equation (\ref{evector}), then the eigenstates are automatically in the Schmidt form \cite{nch}, where the basis states $\{|M-n,n\rangle_{12}\}$ now act as our Schmidt basis.  From this observation is is straightforward to show that $S_{ent}$ equals  twice the entropy of the probability distribution $|c^M_n(E)|^2$, i.e
\begin{equation}
\label{Hent}
S_{ent}=2S\left(|c^M_n|^2\right)=-2\sum_n{|c^M_n|^2\log(|c^M_n|^2)},
\end{equation}
where the logarithms are taken to the base 2.  We can now evaluate $S_{ent}$ as a function of the total number of photons $M$.  Figure \ref{fig3} shows $S_{ent}$ plotted against $M$, with the energy $E$ either fixed or set equal to the largest possible value given the choice of $M$.  We see that as $M$ increases so does $S_{ent}$.  The fact that the entanglement increases with $M$ is not surprising as the dimensions of our systems Hilbert space are $M+1$.  Increasing the number of photons thus gives us more terms in the Schmidt decomposition of the eigenstates $|E^M\rangle_{12}$.  We can also see that the zero eigenstate, i.e. the state $|E^M\rangle_{12}$, with $E=0$, is less entangled than the states with $E=1$ or $E=2$.  By examining equation (\ref{3term}) is can be seen that for $E=0$, we would require that every term $c^M_n$ equal zero for $n$ odd.  The zero eigenvector thus has roughly half the number of Schmidt terms as the other eigenvectors do.  Consequently, the zero eigenvector is not as entangled as the eigenvectors that correspond to $E\ne0$.

\begin{figure}
\center{\includegraphics[width=9cm, height=!]
{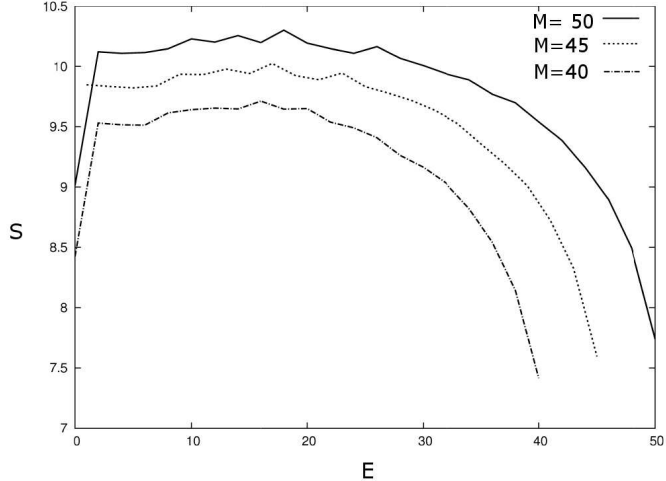}}
\caption{Entanglement, $S_{ent}$, plotted against the energy $E$, for different values of the photon number $M$.}
\label{fig4}
\end{figure}

It is interesting to see how the entanglement changes as we vary $E$, with the total number of photons now fixed.  
Figure \ref{fig4} shows $S_{ent}$ plotted against the eigen-energy, $E$, for several different values of $M$.  It can be seen that when $E$ is large, the entropy decreases.  The fact that there is a general trend for the entanglement to decrease as $E$ increases can be explained by looking at how the probability distribution, $|c^M_n(E)|^2$, changes when $E$ varies.  Figure \ref{fig1} shows that as we decrease $E$ additionally peaks appear in the probability distribution.  This means that the probability becomes more spread out and thus the entropy increases.  Thinking in terms of correlations we see that for $E$ equal to $M|\gamma|$, we have a single maximum in the probability distribution $|c^M_n|^2$.  The Schmidt basis states $|M-n,n\rangle_{12}$ will only have coefficients with large values for the states with $n$ close to $M/2$.  When $E$ is decreased we obtain several maxima in the probability distribution $|c^M_n(E)|^2$ and thus we have more Schmidt basis states with larger coefficients and hence the entanglement increases.

An interesting feature of figure \ref{fig4} is that there are points where the entanglement increase slightly as $E$ increases.  One obvious example of this is when $M$ is even and $E$ increases from zero to $2$ ($\gamma=1$ in figure \ref{fig4}).  In this case the increase is due to the anomalous nature of the zero energy eigenvector, i.e it has roughly half the number of nonzero Schmidt coefficients as the other energy eigenvectors.  The other instances when the entanglement increase occur due to the fact that $P^M_n(E)$ can sometimes broaden when $E$ increases.  Sometimes this broadening will compensate for the loss of peaks in $P^M_n(E)$ and thus the probability distribution is more spread out, which leads to a slight increase in the entropy.

\section{Nonlinear Hamiltonians and conditional state swapping}
\label{sec4} In section \ref{sec2} we obtained the eigenvalues and
eigenvectors corresponding to the quadratic Hamiltonian (\ref{bsh}).
These results can now be used to determine the properties of a
certain class of nonlinear Hamiltonians, which are constructed from
$\hat H_0$ [an example of this is given in equation
(\ref{nonlinear})].

Let $f(y)$ be a polynomial in the variable $y$, it is clear that
$[\hat H_0,f(\hat H_0)]=0$, and thus $f(\hat
H_0)|E^M\rangle_{12}=f(E)|E^M\rangle_{12}$.  Similarly, the operator
$\hat n=\hat n_1+\hat n_2$, obeys $[\hat n,\hat H_0]=0$ and thus
$f(\hat n)|E^M\rangle_{12}=f(M)|E^M\rangle_{12}$.  From this we see
that the general Hamiltonian
\begin{equation}
\label{genh}
\hat H=\sum_k{\omega_k\hat n^k+\alpha_k(\hat H_0)^k},
\end{equation}
can be diagonalised using the eigenvectors $\{|E^M\rangle_{12}\}$.
We should note that the spectrum of equation (\ref{genh}) will
generally be degenerate and thus we can construct several different
sets of eigenvectors for the Hamiltonian.  One important practical
consideration for our nonlinear Hamiltonians is that the magnitude
of the terms $\alpha_k$ should decrease as $k$ increases.  In
practice, if we want the higher order terms in our Hamiltonian to
give an non-negligible contribution to the evolution of our system,
we would have to increase the intensity of our light beams.  This
would correspond to $M$ being large.  Even when we increase the
intensity of the optical beams that enter the nonlinear media, it
will still be difficult to excite higher order terms.  For this
reason we shall look only at examples were $\hat H$ is at most
quadratic in $\hat H_0$, i.e. Hamiltonian with a form  given in
equation (\ref{nonlinear}).

Using our previous results we can easily determine the time
evolution of a system described by a Hamiltonian of the form given
in equation (\ref{genh}).  If at time $t=0$ the system is prepared
in the state $|\psi_0\rangle_{12}$, then at a later time $t$, it
will be in the state $|\psi_t\rangle_{12}=\exp(-i\hat
Ht)|\psi_0\rangle_{12}$, where $\hbar=1$.  The problem of
determining the evolution thus reduces to expressing the state
$|\psi_0\rangle_{12}$ in terms of the basis $\{|E^M\rangle_{12}\}$.
In quantum optics, states are often expressed in terms of the Fock
basis.  It is thus essential to be able to express a given Fock
state in terms of the energy eigenvectors.  This can be achieved
using the completeness relation, $\sum_k{|E_k\rangle\langle
E_k|}=\hat 1$, which enables us to obtain that
\begin{equation}
\label{nstates} |M-n,n\rangle_{12}=\sum_k{_{12}\langle
E_k|M-n,n\rangle_{12}|E_k\rangle_{12}}=e^{ing}\sum_k{c^M_n(E_k)|E_k\rangle_{12}}.
\end{equation}

We now give examples of how the preceding theory can be used to
construct Hamiltonians that perform specific tasks. In order to
simplify the following calculations we set $\gamma$ real.  Suppose
we have the Fock state $|M,0\rangle_{12}$, which we want to
transform to the state $|0,M\rangle_{12}$, i.e. we swap the two
modes.  From equation (\ref{nstates}) it is clear that if we want to
perform this task, then we must change the coefficient $c^M_0(E_k)$
to $c^M_M(E_K)$.  Using equation (\ref{krawpol}) together with some
simple algebra leads to the result that
\begin{equation}
\label{coeftran} c^M_M(E_k)=(-1)^{M+x_{k}}c^M_0(E_k),
\end{equation}
where $x_{k}=(E_k+M|\gamma|)/(2|\gamma|)$.  This result means that
when $M+x_{k}$ is odd, $c^M_M(E_k)=-C^M_0(E_k)$, while for $M+x_{k}$
even $c^M_M(E_k)=c^M_0(E_k)$.  Suppose that we wish to construct a
Hamiltonian that will peform the swap operation
$|M,0\rangle_{12}\rightarrow|0,M\rangle_{12}$ for all values of $M$.
There is an infinity of different Hamiltonians, which achieve this
task.  An example of a simple Hamiltonian that performs the swap
operation is
\begin{equation}
\label{lswap}
\hat H=\frac{\pi}{2|\gamma|\tau}\left(3|\gamma|\hat n+\hat H_0\right),
\end{equation}
where $\exp(-i\hat H\tau)|M,0\rangle_{12}=|0,M\rangle_{12}$.  This Hamiltonian is linear in $\hat H_0$, we could, however, construct alternative nonlinear Hamiltonians that perform the swap operation.  An interesting point to note is that if we had some state $|\psi\rangle=\sum_n{\xi_n|n\rangle}$, then the Hamiltonian described in equation (\ref{lswap}) can be used to enact the transformation $|\psi\rangle_1|0\rangle_2\rightarrow|0\rangle_1|\psi\rangle_2$, i.e. $\exp(-i\hat H\tau)|\psi\rangle_1|0\rangle_2=|0\rangle_1|\psi\rangle_2$.

In the previous example the swap operation was insensitive to the total number of photons.  It is, however, possible to design Hamiltonians that will perform a swap operation only if the number of photons is either even or odd.  Suppose now that we want the swap operation to occur only if the total number of photons is even, i.e $|M,0\rangle_{12}\rightarrow|0,M\rangle_{12}$ only when $M$ is even.  Let $\epsilon^M_x$ denote the energy eigenvalue of our nonlinear Hamiltonian, corresponding to the eigenvector $|E^M_x\rangle_{12}$.  In order to perform the swap at time $t=\tau$ we require that $\exp(-i\epsilon^M_x\tau)=(-1)^{M+x}$, but only when $M$ is even.  
A Hamiltonian that achieves this is 
\begin{equation}
\label{evenswap}
\hat H=\frac{\pi}{|\gamma|\tau}\left(|\gamma|\hat n+\frac{|\gamma|}{4}\hat n^2+\frac{1}{4|\gamma|}\hat H_0^2\right).
\end{equation}
One can easily verify that 
\begin{equation}
\label{eventran}
\exp(-i\hat H\tau)|M,0\rangle_{12}=\begin{cases}
|0,M\rangle_{12},\;\text{for $M$ even},\\
i|M,0\rangle_{12},\;\text{for $M$ odd}.
\end{cases}
\end{equation}
The Hamiltonian (\ref{evenswap}) thus has the property that for a
state $|\psi\rangle$, the transformation
$|\psi\rangle_1|0\rangle_2\rightarrow|0\rangle_1|\psi\rangle_2$ can
only be realized if $|\psi\rangle$ can be expressed as a
superposition of Fock states consisting of an even number of
photons.

We have described how the Hamiltonian given in equation
(\ref{evenswap}), can be used to perform a conditional swap
operation.  The Hamiltonian can also be used to perform other useful
tasks, such as preparing two optical  cat states form a single
coherent state.  An optical cat is a state of the form
$K(|\alpha\rangle+|-\alpha\rangle)$, where $K$ is a normalisation
constant and $|\pm\alpha\rangle$ are coherent states \cite{loudon}.
The significance of these states can be seen by recalling that when
$|\alpha|$ is large, coherent state behave like a classical
monochromatic optical fields.  Cat states are thus a superposition
of two classical objects\footnote{The name `cat state' is a
reference to the thought experiment developed by Schr\"{o}dinger
\cite{schrodinger}.}.  The Fock state representation of a coherent
state is \cite{Klauder}
\begin{equation}
|\alpha\rangle=\exp\left(-\frac{|\alpha|^2}{2}\right)\sum^{\infty}_{n=0}{\frac{\alpha^n}{\sqrt{n!}}|n\rangle},
\end{equation}
where $\alpha$ is a complex number.  If we take our initial state to be $|\alpha\rangle_1|0\rangle_2$, then using the Hamiltonian (\ref{evenswap}) will yield the following evolution
\begin{eqnarray}
\label{cat}
\exp(-i\hat H\tau)|\alpha\rangle_1|0\rangle_2=|0\rangle_1\left(\frac{|\alpha\rangle_2+|-\alpha\rangle_2}{2}\right)+i\left(\frac{|\alpha\rangle_1-|-\alpha\rangle_1}{2}\right)|0\rangle_2.
\end{eqnarray}
We have thus obtained a superposition of two optical cat states.  Theoretical schemes for obtaining a superposition of two optical cat states have been proposed previously \cite{milburn, tombesi, yurke2}.  In each of the earlier schemes, the Hamiltonians that were used were different from that given in equation (\ref{evenswap}).

As we have shown, it is possible to construct Hamiltonians that perform a swap operation conditioned on whether the number of photons is even.  It is also interesting to consider a swap operation conditioned on other factors.  For instance, we may want to not swap the state if it has a particular number of photons, while allowing the swap to occur if the number of photons differs by one.  We now give the form of a Hamiltonian that performs this task.  As before, let $\epsilon^M_x$ denote the energy of the nonlinear Hamiltonian.  We require that $\exp(-i \epsilon^M_x\tau)=(-1)^{M+x}$ if $M=N\pm 1$, but that $\exp(-i\epsilon^M_x\tau)\ne(-1)^{M+x}$ for $M=N$.  It can easily be verified that the following Hamiltonian satisfies the stated conditions
\begin{equation}
\label{pswap}
\hat H=\frac{\pi}{2|\gamma|\tau}\left(3|\gamma|\hat n^2+\hat n\hat H_0-3N|\gamma|\hat n-N\hat H_0\right).
\end{equation}
This leads to the following dynamics, $\exp(-i\hat H\tau)|N,0\rangle_{12}=|N,0\rangle_{12}$, while $\exp(-i\hat H\tau)|N\pm1,0\rangle_{12}=|0,N\pm1\rangle_{12}$.  It is straightforward to modify the Hamiltonian given in equation (\ref{pswap}) for situations where we want to swap states other than $|N\pm1,0\rangle_{12}$.  An example of this is when we want to swap states such as $|N\pm2,0\rangle_{12}$.  To achieve this we can simply multiply equation (\ref{pswap}) by a factor of $1/2$, i.e. we take our Hamiltonian to be $\hat H/2$.  We shall now show how one can use the Hamiltonian (\ref{pswap}) together with our previous nonlinear Hamltonians, so as to discriminate between photon number states.

Suppose we have four modes that are initially prepared in the state
\begin{equation}
\label{nooo}
|M\rangle_1|0\rangle_2|0\rangle_3|0\rangle_4,
\end{equation}
where $M\le 4$.  An important problem to address is whether we can determining the value of $M$.  The majority of currently available photon detectors are not capable of reliably discriminating between different photon number states.  For this reason it would be interesting if we could act on the state (\ref{nooo}), so that the final position of the mode that is not in the vacuum is dependent on the number of photons.  In particular, we could arrange for the state to evolve such that the $M$-th mode contains $M$ photons, while the other modes are in the vacuum state.  We shall now outline a scheme that enables one to performs this task.

At time $t=0$ our system is prepared in the state (\ref{nooo}).  We arrange for modes 1 and 2 to interact in a manner governed by the Hamiltonian (\ref{evenswap}).  This means that at time $\tau$ the photons will have been transferred to the second modes if $M$ is even, or remain in the first mode for $M$ odd.  We now arrange for the first and third modes to be coupled via a Hamiltonian of the form $\hat H/2$, where $\hat H$ is given by equation (\ref{pswap}) and where the $N$ in equation (\ref{pswap}) equals 1.  This will result in the first and third mode being swapped when we have three photon.  At the same time we shall also couple the second and fourth modes using a Hamiltonian of the form $\hat H/2$, where $\hat H$ is given by equation (\ref{pswap}), with $N=2$.  This will result in the second and fourth modes being swapped if the number of photons is four.  The effect of coupling the modes in the manner that we have described is that our system will be left in the state
\begin{eqnarray}
i|1\rangle_1|0\rangle_2|0\rangle_3|0\rangle_4,\;\text{for $M$=1},\nonumber\\
|0\rangle_1|2\rangle_2|0\rangle_3|0\rangle_4,\;\text{for $M$=2},\nonumber\\
|0\rangle_1|0\rangle_2|3\rangle_3|0\rangle_4,\;\text{for $M$=3},\nonumber\\
|0\rangle_1|0\rangle_2|0\rangle_3|4\rangle_4,\;\text{for $M=4$}.
\end{eqnarray}
We could now put photon detectors at each mode so as to determine which mode contains photons.  This in turn would allow us to determine the number of photons that we have.  Alternatively we could perform further operations on the individual modes.  In addition to this, the above scheme can be used to prepare multiphoton entangled states, provided we can prepare a superposition of a finite number of Fock states, in a single modes.  For example, we could transform the separable state $(a|1\rangle+b|2\rangle)|0\rangle$ into the entangled state $a|10\rangle+b|02\rangle$.  The various procedures that we have outlined could find applications in quantum information processing and quantum control.

\section{Dynamics of the nonlinear Hamiltonians}
\label{sec5}
In the previous section we constructed nonlinear Hamiltonians that allowed use to perform certain types of operations on the two coupled modes.  In this section we will investigate further the dynamics induced by the three Hamiltonians given by (\ref{lswap}), (\ref{evenswap}) and (\ref{pswap}).  A natural quantity to study is the number of photons in each mode.  For this reason we shall investigate how the expectation value for the number of photons in the second mode changes, i.e. $\langle\hat n_2\rangle$.  
The initial state of our system will be taken to be $|M,0\rangle_{12}$, and thus $\langle\hat n_2\rangle=0$ at time zero.

\begin{figure}
\center{\includegraphics[width=9cm, height=!]
{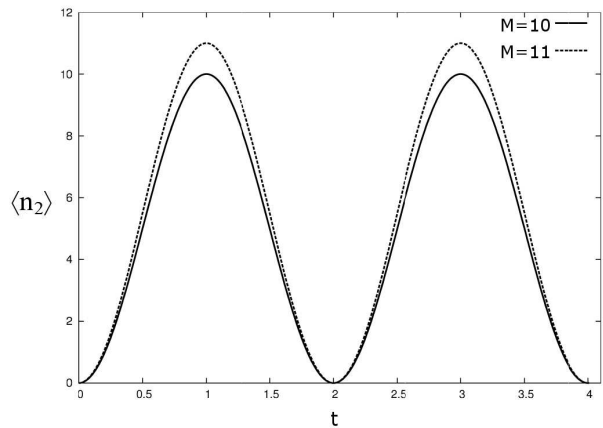}}
\caption{A plot of the expectation value $\langle\hat n_2\rangle$, against time, with $\gamma=\tau=1$ and where the initial state is $|M,0\rangle_{12}$.}
\label{oscillate}
\end{figure}

We shall now study the dynamics associated with the quadratic Hamiltonian (\ref{lswap}).  In figure \ref{oscillate} we see the expectation value $\langle\hat n_2\rangle$ plotted as a function of time, for the initial states $|10,0\rangle_{12}$ and $|11,0\rangle_{12}$, with $\gamma=\tau=1$.  The expectation value oscillates with period of two (i.e. $2\tau$), the period is thus independent of the number of photons.  Another interesting feature of figure \ref{oscillate} is that at time $t=1$, $\langle\hat n_2\rangle=M$, which is consistent with the state having been transformed to $|0,M\rangle_{12}$ at that time.  The oscillatory dynamics shown in figure \ref{oscillate} is also observed for other values of $M$, and does not depend on whether $M$ is even or odd.  One final interesting feature of the Hamiltonian (\ref{lswap}) is that we obtain oscillations for different choices of the initial state.  For example, if the state was initially $|M-n,n\rangle$, then we would still observe oscillations of $\langle\hat n_2\rangle$, with period $2\tau$.  The minimum and maximum values of $\langle\hat n_2\rangle$ would, however, respectively change to $\langle\hat n_2\rangle=n$ at $t=0$ and $\langle\hat n_2\rangle=M-n$ at $t=\tau$.  Such behaviour is consistent with the Hamiltonian (\ref{lswap}) being able to swap the two modes.

\begin{figure}
\center{\includegraphics[width=9cm, height=!]
{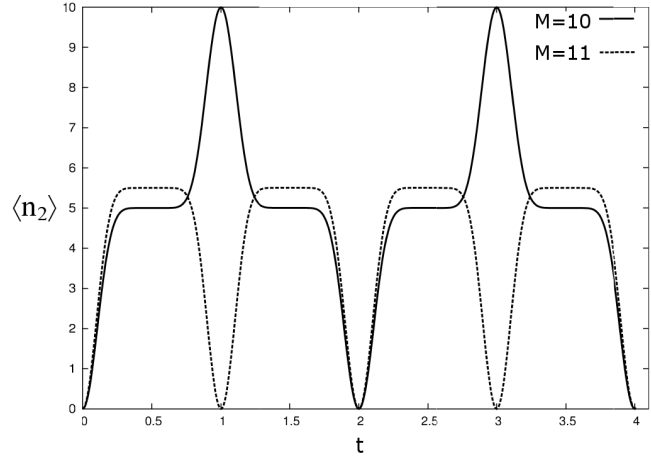}}
\caption{A plot of the expectation value $\langle\hat n_2\rangle$, against time, with $\gamma=\tau=1$ and where the initial state is $|M,0\rangle_{12}$.  It can be seen that for $M=11$ the Hamiltonian acts to trap some of the energy in the first mode, while for $M=10$ the energy is fully swapped between the two modes.}
\label{selective1}
\end{figure}

The Hamiltonian (\ref{evenswap}) was designed to perform the transformation described in equation (\ref{eventran}).
This selective transformation has similarities to the behaviour of a two mode nonlinear directional
coupler \cite{jensen,chefles}.  This is an optical device which consists of two optical fibers that are coupled in
such a way that the dynamics depends on the intensity of the light \cite{chefles}.  One obvious difference between
these two nonlinear systems is that the conditional dynamics associated with the Hamiltonian (\ref{evenswap}),
dependent on whether the number of photons is even or odd and not on how many photons there are.  Figure \ref{selective1} shows a plot of the expectation value $\langle\hat n_2\rangle$, against the time, for the initial states $|10,0\rangle_{12}$ and $|11,0\rangle_{12}$.  We see that the dynamics is very different for the two different initial states.  In particular, $\langle\hat n_2\rangle$ never exceeds the value 5.5, for the case when $M=11$.  This is in contrast to the situation for $M=10$, where we find that $\langle\hat n_2\rangle=10$ at $t=1$ (i.e. $t=\tau$).  In both cases we see that the behaviour of $\langle\hat n_2\rangle$ is periodic, with period $t=2\tau=2$.  When $M=10$, one can observe that the expectation value $\langle\hat n_2\rangle$ temporarily plateaus at the value of 5, which is half the total number of photons.  

\begin{figure}
\center{\includegraphics[width=9cm, height=!]
{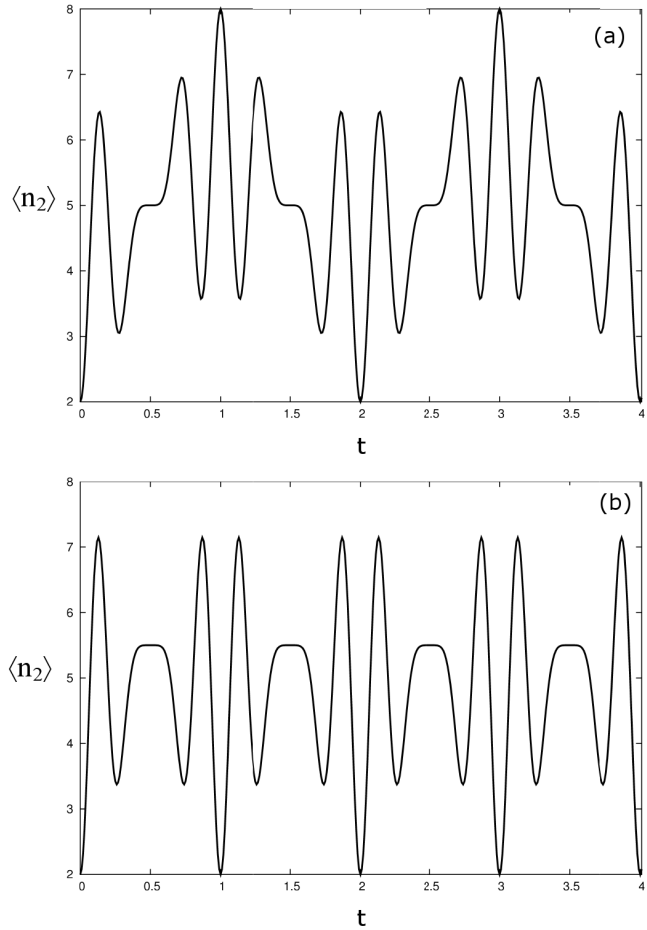}}
\caption{A plot of the expectation value $\langle\hat n_2\rangle$, against time, with $\gamma=\tau=1$.  In plot (a) the initial state is $|8,2\rangle_{12}$, while in plot (b) the initial state is $|9,2\rangle_{12}$.}
\label{selective2}
\end{figure}

It is interesting to investigate the dynamics of the Hamiltonian (\ref{eventran}) for initial states, which have photons in both modes, i.e $|M-n,n\rangle_{12}$, with $n\ne 0$.  Figure (\ref{selective2}) show $\langle\hat n_2\rangle$ plotted against time, for the initial states $|8,2\rangle_{12}$ and $|9,2\rangle_{12}$.  In both cases the minimum of $\langle\hat n_2\rangle$ is now 2.  For the case when $M=10$, we see that at $t=1$, $\langle\hat n_2\rangle=8$, which is consistent with the state undergoing the transformation $|8,2\rangle_{12}\rightarrow|2,8\rangle_{12}$.  When the initial state is $|9,2\rangle_{12}$ and thus $M=11$,  $\langle\hat n_2\rangle$ is always less than 9, and thus there is never a point where we transfer the 9 photons that were initially in the first mode, to the second mode.

The Hamiltonian (\ref{pswap}) was designed so that it would perform the transformation $|N\pm1,0\rangle_{12}\rightarrow|0,N\pm1\rangle_{12}$, while leaving a state with exactly $N$ photons unchanged.  It is clear that if our system is initially prepared in the state $|N,0\rangle_{12}$, then the system will stay in that state.  In fact, any state with exactly $N$ photons will remain unchanged under the action of (\ref{pswap}). 
If our system has $N\pm1$ photons then the dynamics of the Hamiltonian (\ref{pswap}) will be identical to that of (\ref{lswap}).

\section{Conclusions}
\label{conc}
We have investigated a class of nonlinear Hamiltonians.  These  Hamiltonians were constructed from a simple beam splitter
 Hamiltonian, equation (\ref{bsh}), and from the photon number operator $\hat{n} $.  Due to the pivotal role played
 by the beam
 splitter Hamiltonian, (\ref{bsh}), we investigated its properties thoroughly.  The eigenvalues and eigenvectors
 of equation (\ref{bsh}) were obtained using the theory of orthogonal polynomials.  In particular, it was shown that
 the eigenvalue problem associated with equation (\ref{bsh}) was equivalent to solving the three term recurrence relation
 associated with a class of orthogonal polynomials known as Krawtchouk polynomials.  
We were thus able to show that the dynamics of the beam splitter Hamiltonian and the class of nonlinear Hamiltonians,
are exactly solvable using Krawtchouk polynomials.  

Nonlinear Hamiltonians were then constructed from the beam splitter
Hamiltonians and the total photon number operator.   The dynamics of
these Hamiltonians enabled novel tasks such as state swapping and
conditional state swapping, to be performed. The formalism that we
have introduced and the Hamiltonians that we have studied, have
applications in quantum optics and quantum information.  In
particular, they can be used in the areas of state preparation and
in quantum control.  For example, we have shown how the nonlinear
Hamiltonian (\ref{evenswap}) can be used to prepare a superposition
of Schr\"{o}dinger cats states for a coherent state.  In addition to
this we have shown how we can control several modes so as to sort
photons into one particular mode, where the mode that there are
transferred to depends on the total number of photons. This was
achieved by turning on couplings between two different modes, where
the interactions were described using the nonlinear Hamiltonians
(\ref{evenswap}) and (\ref{pswap}).  While we have discussed this
work in terms of optics and photons, the results would apply to any
system of bosonic particles.


\section*{Acknowledgements}
We acknowledge financial support from grants MSM6840770039 and MSMT
LC06002 of the Czech Republic. The authors are obliged  prof. A.
Odzijewicz for helpful comments and remarks.


\begin{thebibliography}{00}
\bibitem{para} P. G. Kwiat, K. Mattle, H. Weinfurter, A. Zeilinger, A. V. Sergienko and Y. Shih {Phys. Rev. Lett.}, {\bf 1995}, 75, 4337-4341.
\bibitem{jensen} S. M. Jensen, {IEEE J. Quantum Elect.}, {\bf 1982}, EQ-18, 1580-1583.
\bibitem{chefles} A. Chefles and S. M. Barnett, {J. Mod. Opt.}, {\bf 1996}, 43, 709-727.
\bibitem{milburn} G. J. Milburn and C. A. Holmes, Phys. Rev. Lett. {\bf 1986}, 56, 2237-2240.
\bibitem{yurke} B. Yurke and D. Stoler, Phys. Rev. Lett. {\bf 1986}, 57, 13-16.
\bibitem{tombesi} A. Mecozzi and P. Tombesi, Phys. Rev. Lett. {\bf 1987}, 58, 1055-1058.
\bibitem{yurke2} B. Yurke and D. Stoler, Phys. Rev. A {\bf 1987}, 35, 4846-4849.
\bibitem{askey} R. Koekoek, R. F. Swarttouw, The Askey-scheme of hypergeometric orthogonal polynomials and its q-analogue, Report DU 98-17, TUDelft, 1998; http://aw.twi.tudelft.nl/~koekoek/askey.html
\bibitem{Goce2} M. Horowski, G. Chadzitaskos, A. Odzijewicz and A. Tereszkiewicz {J. Phys. A}, {\bf 2004}, 37, 6115-6128.
\bibitem{Goce} A. Odzijewicz, M. Horowski and A. Tereszkiewicz {J. Phys. A}{\bf 2001}, 34, 4353-4376.
\bibitem{Horow} M. Horowski, A. Odzijewicz, and A. Tereszkiewicz {Cz. J. Phys.}{\bf 2002}, 52, 1231-1237.
\bibitem{klauder} B. Yurke, S. L. McCall and J. R. Klauder, {Phys. Rev. A},{\bf 1986}, 33, 4033-4054.
\bibitem{campos} R. A. Campos, B. E. A. Saleh and M. C. Teich, {Phys. Rev. A}, {\bf 1989}, 40, 1371-1384.
\bibitem{leonhardt} U. Leonhardt, {Phys. Rev. A}, {\bf 1993} 48, 3265-3277.
\bibitem{orthop} T. S. Chihara, {An introduction to orthogonal polynomials}, Gordon and Breach, New York, 1978.
\bibitem{mathsphys} H. S. Wilf, {Mathematics for Physical Science}, Dover publishing, New York, 1962.
\bibitem{favard} J. Favard, Sur les polynomes de Tchebicheff, C. R. Acad. Sci. Paris Sér. I Math.{\bf 1935}, 200, 2052-2053.
\bibitem{nch} M. A. Nielsen and I. L. Chuang, Quantum Computation and Quantum Information, Cambridge University Press, Cambridge, 2000.
\bibitem{steve}  S. M. Barnett and S. J. D. Phoenix {Phys. Rev. A} {\bf 1989}, 40, 2404-2409.
\bibitem{steve2} S. M. Barnett and S. J. D. Phoenix {Phys. Rev. A} {\bf 1991}, 44, 535-545.
\bibitem{Loudon} R. Loudon, {The Quantum Theory of Light}, Oxford University Press, Oxford, 2000.
\bibitem{Klauder} J. R. Klauder and  E. C. G. Sudarshan, {Fundamentals
of Quantum Optics,} W. A. Benjamin, Inc., New York, Amsterdam, 1968.
\bibitem{schrodinger} E. Schr\"{o}dinger, {Naturwissenschaften} {\bf 1935}, 23, 807-812.

\end{thebibliography}
\end{document}